\documentclass[11pt,twoside]{article}
\usepackage{asp2010}

\resetcounters

\begin{document}

\title{Visualising Large Datasets in TOPCAT v4}
\author{M.~B.~Taylor}
\affil{H.~H.~Wills Physics Laboratory, University of Bristol, U.K.}

\begin{abstract}
TOPCAT is a widely used desktop application for manipulation of
astronomical catalogues and other tables, which has long provided fast
interactive visualisation features including 1, 2 and 3-d plots, multiple
datasets, linked views, color coding, transparency and more.

In Version 4 a new plotting library has been written from scratch
to deliver new and enhanced visualisation capabilities.
This paper describes some of the considerations in the design and
implementation, particularly in regard to providing comprehensible
interactive visualisation for multi-million point datasets.
\end{abstract}

\section{Introduction}

TOPCAT (Tool for OPerations on Catalogues And Tables)
is a desktop application for manipulating tabular data,
typically source catalogues.
Its aim is to facilitate all the mechanical operations required
when working with tables, including data and metadata browsing,
crossmatching, column calculations, row selections,
and access to Virtual Observatory and other external services,
so that astronomers can concentrate on actually extracting science
from the data.
It is open source,
written in pure Java,
and can comfortably deal with millions of
rows and hundreds of columns on standard desktop or laptop machines.

Sophisticated multi-dimensional visualisation capabilities have
long been part of its capabilities, but a number of requirements
were proving difficult to accomodate into the existing framework,
so version 4, released in March 2013, includes a rewrite from
scratch of the plotting classes to deliver more flexibility,
extensibility, and improved responsiveness and performance for
large datasets.

This paper does not list the new features in detail, but discusses
some of the design considerations and solutions that went into the
development, especially with a view to scaling up to larger datasets.

\section{Variable-Density Plots}

When producing interactive visualisation for large point clouds
at least two issues present themselves.
The first concerns performance: it is most important to provide
fast and responsive rendering of the graphics so that the user can make
many exploratory changes
without having to wait for screen refreshes.
The more fluid the interactive experience, the better it is for
data investigation, and hence scientific results.
Much work has gone into producing efficient code to this end,
though more remains to be done.

However, a question of at least equal importance is how to
present a large dataset to the user in a comprehensible way.
Supposing a collection of ten million points, each with multiple attributes,
how can a user equipped with a million-pixel screen and a human brain
understand the information it embodies, both at the scale of
the whole dataset, and at the scale of individual elements?
Since the datasets we are interested in are typically source
catalogues, each item has importance of its own as well as 
as part of a larger whole
--- one of those rows might be your favourite object,
and you may want to examine its characteristics in detail.

Clearly a single view cannot convey information about each individual
element of a multi-million-row collection.
The capability for the user to navigate interactively between the
large scale (high density regime)
and small scale (low density regime)
is therefore essential.
This in turn raises the question of how to represent the data in
a way that makes sense for both high and low point densities.
Visualisation tools, including earlier versions of TOPCAT,
have typically provided representations suitable for each regime
but lack a single representation that works well for both.
This lack inhibits fluid navigation between large and small scale
views, and also makes it difficult to see both high and low density
regions of a single plot at once.

\begin{figure}[t]
\plotone{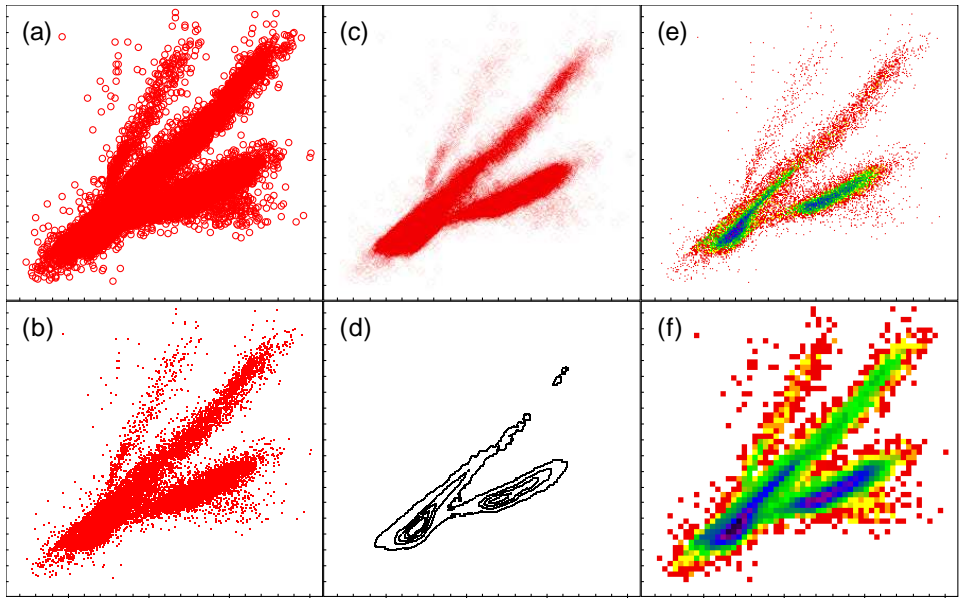}
\caption{Standard graphical representations of a point cloud in 2 dimensions}
\label{fig:O10_f1}
\end{figure}

Figure~\ref{fig:O10_f1} shows some conventional representations for
point cloud data in two dimensions.
In low density regions a conventional scatter plot with medium-sized
markers (a) is most appropriate, but it does not reveal high-density
structure.  For higher density regions smaller markers
work better, down to a single pixel (b), but even this
fails for point densities greater than one per pixel,
and single points are too small to see easily at lower densities.
Plotting markers with configurable transparency (c),
provided in earlier versions of TOPCAT,
can work well for relatively small density variations, but
otherwise tends to lose detail at one or both ends of the density scale.
Contour plots (d) are good for high density but poor at low density.
2-d histograms, also known as density maps,
in which each bin is represented as a square block of pixels
colored according to point count,
can offer a reasonable compromise, but choosing the bin size
presents problems: small bins (e) tend to result in
noisy images and make single points hard to see,
while large bins (f) lose positional resolution by quantising the space.
It is also impossible in density maps to distinguish species
by using different marker shapes.

\begin{figure}[t]
\begin{center}
\includegraphics[height=5.4cm]{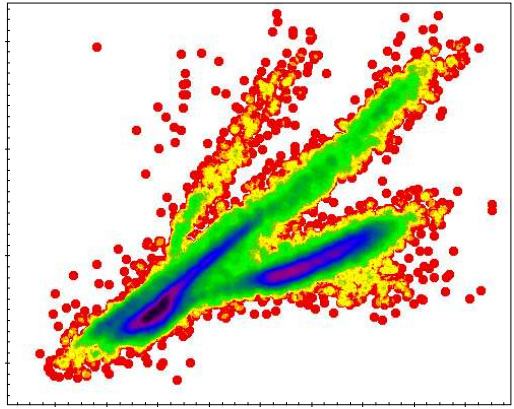}
\end{center}
\caption{Hybrid scatter plot/density map}
\label{fig:O10_f2}
\end{figure}

In TOPCAT v4 we introduce a hybrid scatter plot/density map
(Figure~\ref{fig:O10_f2}).
This is a convolution of a density map with a shaped marker.
A pixel-scale grid is used as for a single-pixel density map,
but for each data point, instead of incrementing the count in a single bin,
the count is incremented in all bins covered by the shaped marker
assigned to the data species.
Pixels are then colored by count according to a configurable color map.
Another way to consider this is as a scatter plot with non-standard
compositing.

The great benefit of this representation is that it
transitions smoothly between a scatter plot at low densities
and a (smoothed) density map at high densities.
Not only can this simultaneously reveal high-density structure
and outlier positions,
but it also works well for low and high magnification of the same data.
This latter is particularly useful as it means the user can
zoom in and out interactively between large scale structure
and single objects without having to change between plot types.

While this works well for single-species plots in two or three dimensions,
it should be noted that
it lends itself less well to multiple overplotted datasets,
since if color codes for density, it cannot be used so easily to identify
different species.
Various options are provided to address this, including
use of different shaped markers and per-species color maps that code
density by the {\em Value\/} component of the color HSV triple.

\section{3-Dimensional Navigation}

Interactive navigation in two dimensions is straightforward,
since a mouse can be used to ``grab'' and drag the plot around,
and zooming can be centered around the mouse position.
An intuitive user interface for three dimensional navigation is
harder to define however, in particular because the 2-d mouse position
corresponds to a line of sight rather than a point, so does not define an
unambiguous center for zooming operations.

This degeneracy is broken by taking the mouse position to indicate for
zooming purposes the ``center of mass'' of the data points
visible along the line of sight it denotes.
For typical point clouds this line of sight center of mass rule
works well; pointing at a high-density cluster usually yields
the center of the cluster, and pointing at an isolated object
along a low-density line of sight yields that object's position.
It is not possible to navigate in this way to the center of a void,
however, this is often not required.

\section{Configuration Architecture}

Since target datasets vary widely, it is essential to provide many
configuration options for visualisation; a few examples are
marker color, size and shape,
axis ranges and annotations,
color maps and scaling,
contour levels and smoothing,
legend appearance and placement.
A typical plot is controlled by several tens of options
alongside the actual data coordinate values.
This complexity presents challenges in both user interface design and
plotting implementation.

For specifying many options a large user interface is unavoidable,
presenting a potentially serious usability issue.
This is mitigated by ensuring that all options have default values
that together give a reasonable plot.
In particular, when a plotting window is first opened it always
tries to display some plausible plot rather than a blank view.
The user interface then provides a range of configuration controls
grouped by function
that can be adjusted to produce the plot that the user wants to see.

From the implementation point of view, each configuration option
is represented by a ``ConfigKey'' object.
Each key can supply user metadata (name, description),
value type,
a sensible default value,
a GUI component for specifying values,
and methods for mapping between typed values and string representations.
Each plot type (e.g.\ 2D, 3D, Sky)
and each data layer type
(e.g.\ scatter plot, error bars, contours, analytic function)
can report the keys that together specify its configuration.
Different components of the plotting system then make use of these to
build the user interface and gather configuration information
without hard-coded knowledge of each plot and layer type.
In TOPCAT, the plot window establishes which plot type and layers are
in use, interrogates them for their ConfigKeys,
and builds the user interface by stacking the relevant GUI components.
When the user interacts with any of these, the plotting system
establishes the value for each one, bundles the values together,
and passes them to the plotting classes to generate a drawing on the screen.
Since values can be mapped to strings, the configuration may be
represented as a map of string-string key-value pairs,
making it possible to drive the plotting externally, for instance
from a command-line interface or via inter-process communication.
Finally, the user documentation for each plot type can be generated
programmatically at documentation build time by interrogating
each key for its user metadata.
The result is that new configuration options can be introduced
easily by making only localised changes to plot type or layer type code.

\section{Current and Future Status}

TOPCAT v4.0 provides flexible, extensible, and high performance visualisation.
In its existing form as a general rule
it performs well for datasets up to around a
million rows and tolerably for those in the ten million range,
though users have reported plots with up to 300 million rows.
It is hoped to push these numbers up by around an order of magnitude
by multithreading and other optimisations within the next year or two.
More plot types and data layer types will also be introduced.

\end{document}